\documentclass[]{spie}  

\newcommand{\etal}{{\it et al.}}

\usepackage{amsmath,amsfonts,amssymb}
\usepackage{graphicx}
\usepackage[colorlinks=true, allcolors=blue]{hyperref}

\title{Cold optical design for the Large Aperture Simons Observatory telescope.}

\author[a]{S.R. Dicker}
\author[b]{P.A. Gallardo}
\author[c]{J.E Gudmundsson}
\author[d]{P.D. Mauskopf}
\author[e]{A. Ali}
\author[e]{P.C. Ashton}
\author[a]{G. Coppi}
\author[a]{M.J. Devlin} 
\author[g]{N. Galitzki}
\author[h]{S.P. Ho}
\author[e]{C.A. Hill}
\author[i]{J. Hubmayr}
\author[e]{B. Keating} 
\author[e]{A.T. Lee} 
\author[a]{M. Limon}
\author[j]{F. Matsuda} 
\author[k]{J. McMahon}
\author[b]{M.D. Niemack}
\author[a]{J.L. Orlowski-Scherer}
\author[l]{L. Piccirillo}
\author[o]{M. Salatino}
\author[k]{S.M. Simon}
\author[h]{S.T. Staggs}
\author[m]{R. Thornton}
\author[i]{J.N. Ullom}
\author[b]{E.M. Vavagiakis}
\author[n]{E.J. Wollack}
\author[a]{Z. Xu}
\author[a]{N. Zhu}
\affil[a]{Department of Physics, University of Pennsylvania, Philadelphia, PA, USA}
\affil[b]{Department of Physics, Cornell University, Ithaca, NY, USA}
\affil[c]{The Oskar Klein Centre for Cosmoparticle Physics, Department of Physics, Stockholm University, Stockholm, Sweden}
\affil[d]{School of Earth and Space Exploration and Department of Physics, Arizona State University, Arizona, USA}
\affil[e]{Department of Physics, University of California, Berkeley, Berkeley, California, USA}
\affil[g]{Department of Physics, UCSD, La Jolla, CA, USA}
\affil[h]{Department of Physics, Princeton University, Princeton, NJ, USA}
\affil[i]{Quantum Sensors Group, NIST, Boulder, CO, USA}
\affil[j]{The University of Tokyo Institutes for Advanced Study, The University of Tokyo, Kashiwa, Chiba 277-8583, Japan}
\affil[k]{Department of Physics, 450 Church St., Ann Arbor, MI 48109, USA}
\affil[l]{School of Physics and Astronomy, University of Manchester, UK}
\affil[m]{Department of Physics and Engineering, West Chester University of Pennsylvania, West Chester, PA, USA}
\affil[n]{Goddard Space Flight Center, NASA, Greenbelt, MD, USA}
\affil[o]{AstroParticle and Cosmology laboratory, Paris Diderot University, Paris,75013, France}
\authorinfo{Send correspondence to sdicker@hep.upenn.edu}

\begin{document} 
\maketitle

\begin{abstract}
The Simons Observatory will consist of a single large (6~m diameter) telescope and a number of smaller ($\sim$0.5~m diameter) refracting telescopes designed to measure the polarization of the Cosmic Microwave Background to unprecedented accuracy.  The large aperture telescope is the same design as the CCAT-prime telescope, a modified Crossed Dragone design with a field-of-view of over 7.8 degrees diameter at 90~GHz.  This paper presents an overview of the cold reimaging optics for this telescope and what drove our choice of 350--400~mm diameter silicon lenses in a 2.4~m cryostat over other possibilities.  We will also consider the future expandability of this design to CMB Stage-4 and beyond.  
\end{abstract}


\section{INTRODUCTION}
\label{sec:intro}   

Although the Cosmic Microwave Background (CMB) has already reshaped our knowledge of the contents, origins, and evolution of our universe, the next stage of high precision CMB observations has the potential to revolutionize our understanding.
To this aim, the Simons Observatory (SO) Collaboration is building a series of telescopes.  Located at an altitude of 5200~m in the Atacama Desert, a site well known for excellent visibilities at millimeter wavelengths, these telescopes will observe the sky with polarization-sensitive detectors operating from 27 to 270~GHz, a frequency range chosen to allow the separation of foreground effects and the polarized CMB signal that contains many different signatures of the early Universe.  

To achieve these scientific goals, angular scales between a few arcminutes to well over a degree need to be measured with high fidelity  -- something a single instrument design would struggle to do.  Instead, a large 6~m telescope will be used to measure the small to medium angular scales and a complementary series of $\sim$0.5~m telescopes will be used to recover the larger angular scales.  Sufficient overlap in their coverage of different angular scales will allow accurate cross calibration.  This paper focuses on the optical design of the camera to be used on the large telescope. First, the design of the telescope is described and next, the cold optics designed to be used with it.  An overview of the performance of the chosen cold optical design is given, and possible upgrades to the design are discussed.

\section{THE LARGE APERTURE TELESCOPE}\label{sec:LAT}
The SO Large telescope was designed and is being built with the aim of being able to use it past the lifetime of the SO.  There are some basic requirements in terms of resolution and throughput that had to be met, but at the same time many secondary considerations, affecting cost and schedule, were taken into account in our optical design. These are outlined in the next few sections.
\subsection{Requirements}
\subsubsection{Focal plane size}
State of the art millimeter-wave detectors, used in current experiments, are photon noise limited. Their sensitivity is limited by the random arrival of photons, not by the noise in the detector or readout. To achieve better sensitivity, one must either observe at a site/telescope with a lower background, observe for longer, or with more detectors.  Since current CMB measurements are multi-year campaigns from some of the best millimeter-wave sites on earth, then the only ground-based option is more detectors.  The next generation of CMB instruments will require focal planes containing tens of thousands of detectors.  The initial deployment plans for the SO are for at least 35\,000 detectors on the large aperture instrument and a further 35\,000 on the smaller telescopes -- requiring large fields-of-view (FoV).  To make best use of focal plane space, we plan on using multichroic pixels. Models of foregrounds show that the largest fraction of detectors need to be in the 90~GHz and 150~GHz atmospheric windows, so one requirement is that most of the telescope's focal plane should be usable at these wavelengths.  Possible upgrade paths to a Stage-4 CMB instrument\cite{2017arXiv170602464A}, which would need a focal plane containing over 100\,000 detectors at 90/150~GHz, are highly desirable.  

\subsubsection{Resolution}
Science drivers such as lensing, galaxy clusters, and extragalactic point sources benefit from better than 3$^\prime$ resolution at all frequencies.  At 90~GHz, this requires an aperture size just under 6~m.   Critical to the success of CMB experiments is control of systematic errors such as spillover. For this reason it is advantageous to under-illuminate the mirrors.  Consequently, a 6~m diameter aperture with 5.5~m of it illuminated was baselined.

\subsubsection{Systematic errors}
For further reduction of systematic errors it was decided that the telescope should be off-axis -- Although it is easier to obtain larger fields-of-view with an on-axis telescope, such designs inevitably have higher systematic errors such as sidelobes and higher loading on the detectors due to diffraction at the blocked aperture.  This more than offsets the advantages of a larger focal plane making the off-axis choice more desirable.  Low cross-pol and the ability to easily implement co-moving ground shields that redirect scattered light to the sky are also important requirements.

\subsection{Design}
A number of possibilities were investigated, including off-axis Gregorian telescopes (similar to ACT\cite{ACTtelescope} and SPT\cite{SPTtelescope}), three mirror telescopes, and the Crossed Dragone (CD) design.  As well as meeting the above requirements for FoV and systematics, the designs were also evaluated for their potential cost, compactness, and feasibility.  The CD design is known to have a large FoV and excellent cross polar properties \cite{Tran2008}, but this comes at the price of a much larger secondary mirror. We determined that the Gregorian designs could not meet our FoV requirements. A number of off-axis designs based on 3 mirror astigmatic telescopes where investigated. However, these designs were far less compact with large distances between large mirrors, resulting in a structure that would be difficult to build.  In addition, an extra warm mirror would contribute to detector loading, thus reducing sensitivity per detector, and no compensating increase in FoV over the CD design was found.  As a result, we chose a CD design for the SO large telescope. 

\begin{figure}
\begin{center}
  \includegraphics[height=13cm,angle=0]{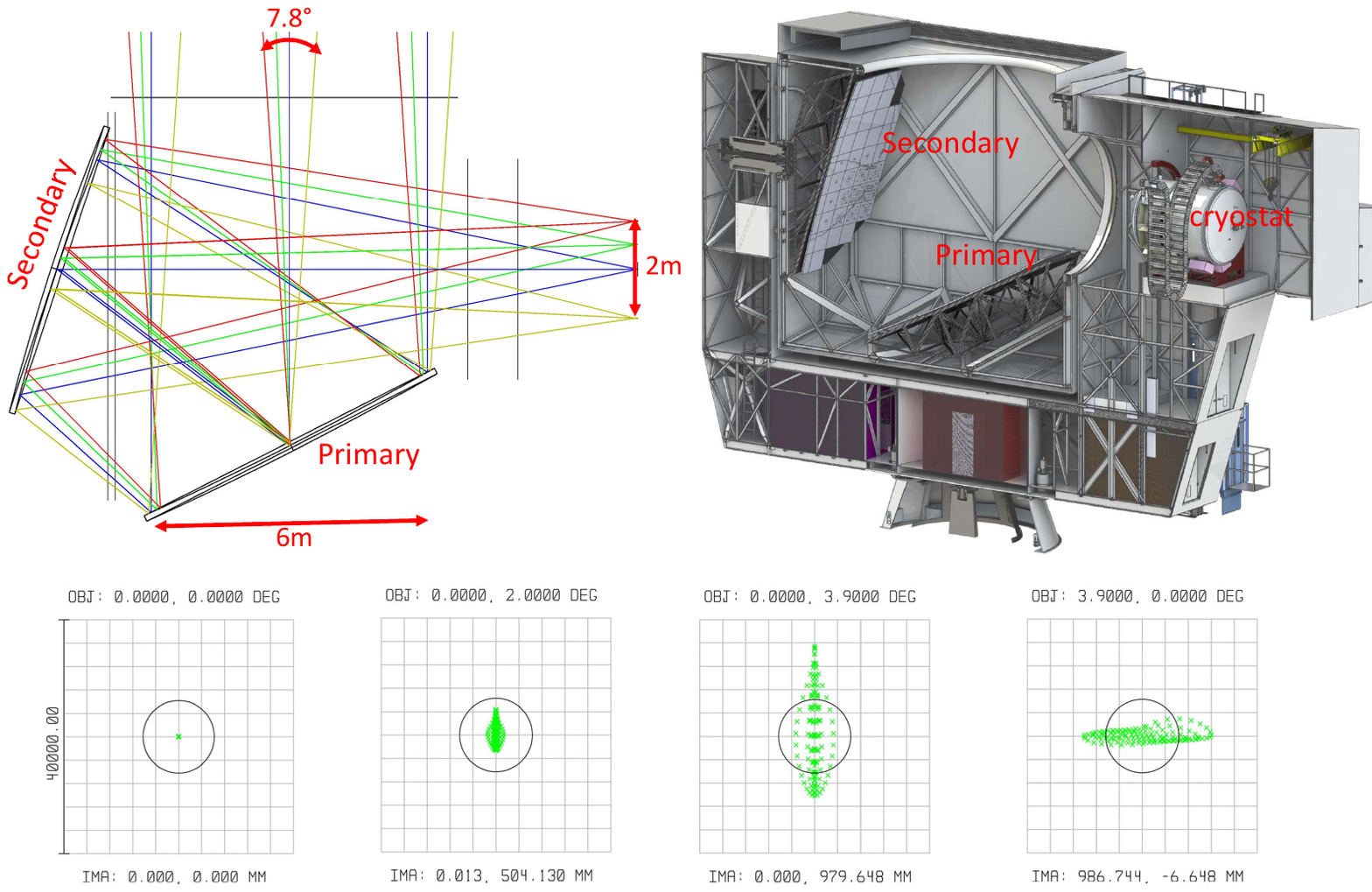}
\end{center}
\caption{\label{telescope} The Simons Observatory/CCAT telescope.  Shown on the top left is a ray trace and on the top right a cross-section through the mechanical design developed by Vertex.  Further details of this design can be found in Parshley \etal\cite{CCATtelescope}.  Along the bottom are spot diagrams with airy disks at 150~GHz. For a regular CD telescope, coma is the dominate aberration in the outer fields however with the extra corrections then the dominate aberration is astigmatism.}
\end{figure}

Having chosen a Crossed-Dragone design, it was realized that our requirements in terms of aperture and frequency range were sufficiently similar to those of the  CCAT-prime project\cite{CCATtelescope2}\cite{Niemack16} so it was agreed to adopt their telescope design.  The classic CD design consists of a parabolic primary and a hyperbolic secondary.  As described in Parshley \etal \cite{CCATtelescope}, it is possible to add canceling spherical terms to each mirror which cancel out comma in off-axis fields.  With these corrections the primary aberration left is astigmatism (Figure~\ref{telescope}).  This comes at the cost of higher field curvature and rotational symmetry being lost from the mirrors. The loss of rotational symmetry means each mirror panel has a different shape but the increase in the diffraction limited field of view outweighed the extra expense.     
The optical design, adopted by both projects, is shown in Figure~\ref{telescope} along with a preliminary design by the contractor selected to built it, Vertex Antennentechnik\footnote{Vertex Antennentechnik GmbH; Baumstr. 46-50; 47198 Duisburg; Germany}. 

\section{COLD OPTICAL DESIGN}
\subsection{Requirements}
The baseline SO detectors\cite{SO_detectors} will be a mixture of dichroic feedhorn and lenslet, antenna coupled Transition Edge Sensor (TES) bolometers cooled to 100~mK by a dilution refrigerator and read out using a $\mu$MUX multiplexing system\cite{umuxReadout}.  Cold optics are needed to reimage the secondary focus of the telescope onto the detector arrays while preserving good image quality and low cross-pol. At the same time, the optical design must provide a cold Lyot stop to prevent the detectors from seeing the warm telescope structure around the mirrors, and have room to put a filter stack capable of, not just defining the band, but also of strongly rejecting out-of-band optical loading, in order to maintain cryogenic performance.  Control of stray light is key. Artifacts such as ghost images can complicate analysis by producing sidelobes that change across the array.  Far more serious is the extra loading scattered light can cause.  The atmospheric loading at the SO site is low. At some frequencies, 2\% of spillover to 300K can reduce the mapping speed by a factor of 1.5\cite{Hill2018}.  Finally, the optics have to be designed with cost, modularity, future expandability, and effective use of the focal plane in mind.

\subsection{Design}
With a proposed focal plane 2~m in diameter, a single vacuum window was ruled out.  Aside from the risk, if it were to be thick enough to withstand one atmosphere, then emission from the window would place an unacceptable load on the detectors.  No single large optical chain, neither lenses nor off-axis mirrors, could reimage the entire focal plane while meeting our requirements for a cold stop, image quality, and scattered light. Having the entirety of our very wide (27~GHz to 270~GHz) band go through the same refracting elements would make for very challenging anti-reflective coatings. For these reasons we break up our total band into 3 sub-bands which we call low frequency (LF) ($\sim$27--60~GHz), mid frequency (MF) ($\sim$90--150~GHz), and ultra-high frequency (UHF) ($\sim$220--270~GHz) with separate reimaging optics for each band.  Taking into account the SO's many different scientific goals and the expected noise in each band, the optimal distribution is to have approximately 13\%,\ 60\%, and 27\% of detectors at LF,\ MF, \& UHF respectively.

Using smaller off-axis mirrors for each band was ruled out since they do not pack well and would make poor use of the telescope focal plane. Instead, cooled refracting optics were chosen.  These have had a proven history on other CMB telescopes\cite{ACTtelescope,SPTtelescope}. As described in Zhu \etal\cite{Zhu10708-79} there are major cost and practical issues in having a cryostat with external dimensions larger than 2.4~m. These include the availability of raw materials and transportation of the cryostat.  Allowing room for internal structure, thermal breaks, readout, \& heat shields of what is a very large cryostat, available room for optics was restricted to a cylinder 205~cm in diameter and a preference to well under 200~cm long (for ease of installation). 

Of the lens materials considered, plastics are by far the least expensive but have too low a refractive index and result in very thick lenses with unacceptable loss.  Higher index materials include silicon and alumina, both of which have a proven track record in CMB experiments.  Wide-band anti-reflective (AR) coating techniques exist for both. With silicon, a multi-layer meta-material can be made by cutting the surface with a dicing saw\cite{siliconAR}, and with alumina, loaded epoxy coatings can be used\cite{AluminaAR}. High purity silicon is readily available in diameters up to 46~cm while alumina lenses can be larger with deployed experiments such as BICEP3\cite{2016SPIE.9914E..0SG} having lenses over 500~mm in diameter. Although both materials are low loss, silicon can be significantly lower\cite{Lamb1996} so silicon will be used where possible and alumina for lenses with physical sizes over 46~cm.

\subsubsection{Optical tube size and spacing}
\begin{figure}
\begin{center}
  \includegraphics[height=10cm]{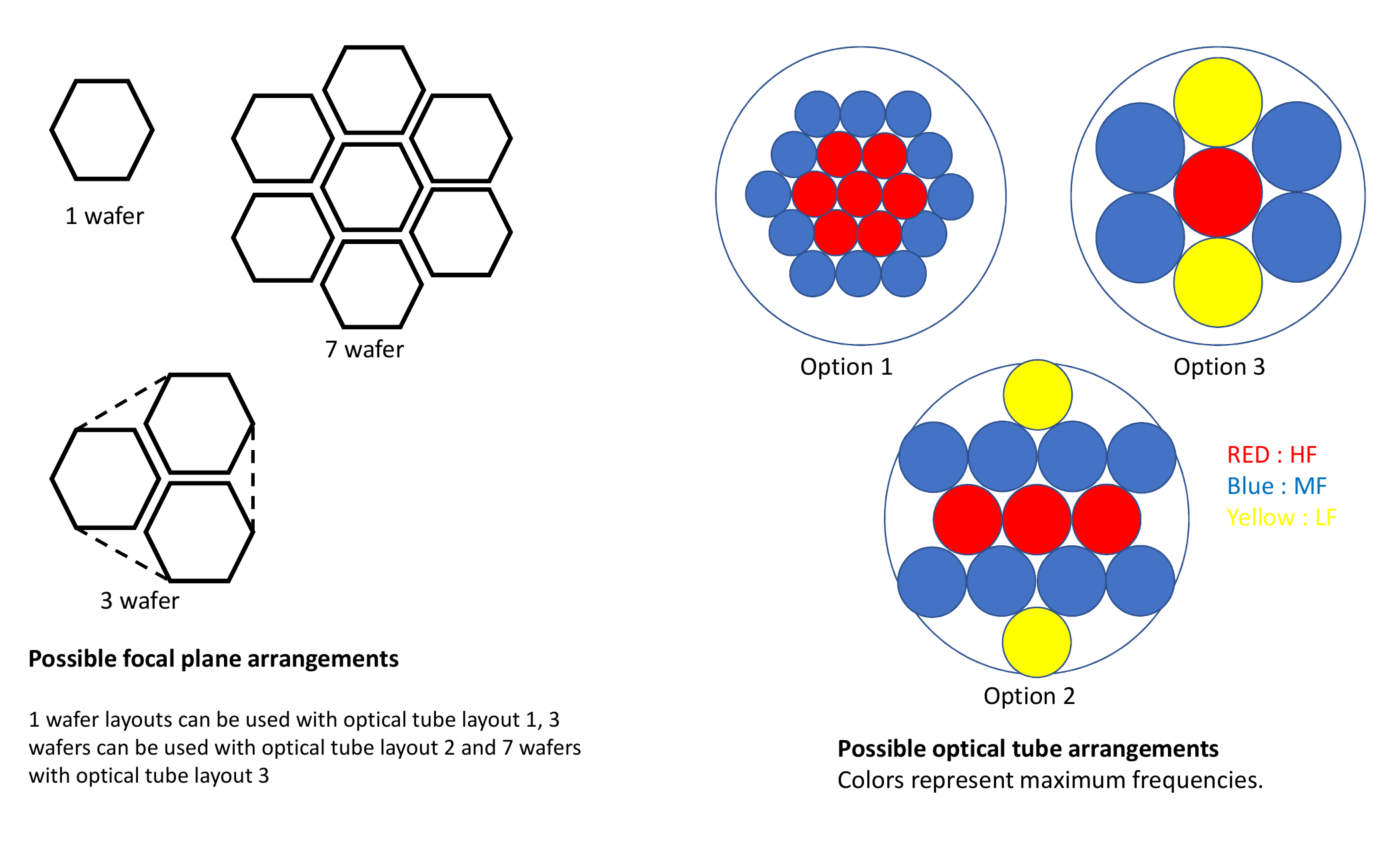}
\end{center}
\caption{\label{fig:arrays}Left: possible layout of focal planes.  Note that the 3 wafer design can be made slightly more efficient by the addition of semi-hexes (shown dashed) at the corners, however this comes at the cost of extra complexity in manufacturing and testing and is not currently the SO baseline plan.  Right: possible layouts for optics tubes within a cryostat (black circle) with sizes that fit 1, 3, and 7 wafers.}
\end{figure}

The SO detectors will be fabricated on wafers 150~mm in diameter\cite{SO_detectors}.  Focal planes are made up of tiling hexagons made from these wafers.  As detector fabrication is expensive, it is cost prohibitive to not use the whole wafer.   Consequently, focal planes come in natural diameters around 150~mm (1 wafer), 275~mm (3 wafers), \& 450~mm (7 wafers)  (Figure~\ref{fig:arrays}).  Around the optically active part of each array, room is needed for; structure to hold it; thermal isolation; and magnetic and thermal shielding.  Crude designs showed a minimum of 25~mm in radius is needed meaning that optical tubes housing 1, 3, \& 7 wafer arrays have a minimum physical size of 200~mm, 375~mm, and 500~mm respectively.

Regardless of what is possible with the optical design, the maximum useful tube size is in part set by the smallest detector that can be fabricated.  Many components of the detectors are of fixed size -- bolometer legs need to be long enough to achieve thermal isolation, inductors and capacitors have minimum sizes, the antenna or feedhorn must be big enough, and room is needed between pixels to get wiring out to the readout.  For SO detectors at MF, this results in minimum pixel sizes of the order of 5~mm\footnote{The exact minimum size depends on frequency and the technology used however on the timescale of SO, significant decrease in pixel size is not expected.} -- or about 30 pixels across a single wafer. Although by using very fast reimaging optics it would be possible for a single wafer to make use of the full FoV of a large optics tube, individual pixels would be widely spaced on the sky, reducing mapping speed.  As shown in Hill \etal\cite{Hill2018}, once photon correlations are taken into account, pixel spacings between 0.25 and $1.2 f \lambda$ are optimum.  The combination of a maximum $1.2 f \lambda$\ pixel spacing and a 5~mm pixel size in our primary science band of 150~GHz gives maximum focal plane usage of 231~mm, 424~mm \& 694~mm for 1 wafer, 3 wafer, and 7 wafer arrays respectively.  Folding in factors such as the maximum filter size available, three different options were compared (Figure~\ref{fig:arrays}); (1) a cryostat filled with 19 tubes spaced by 300~mm containing 1 wafer each, ; (2) 13 tubes spaced by 450~mm containing 3 wafers each; (3) 7 tubes spaced by 600~mm containing 7 wafers each. The pros and cons of each design are described next.

\textbf{Focal plane usage/throughput.}  Like the arrays, an optical tube requires space for mechanical support, thermal isolation, and magnetic shielding.  As this is approximately constant with optics tube size then as a fraction of the area of the focal plane that is optically active, larger tubes can do better.  One exception to this rule is that an integer number of tubes is required so that some optics tube sizes make poor use of the cryogenic volume available.  This turned out to be the case with option 3. A suggestion to enable 100\% focal plane use with small tubes was suggested in Niemack \etal\cite{Niemack16}. Here, the first lens of each tube is used as the vacuum window, allowing light to be focused to a footprint small enough for the physical structure between tubes inside the cryostat.  Due to concerns about the use of lenses as vacuum windows and the alignment of the optics, this concept was considered to be high risk and beyond the scope of the SO project.  Within the constraints of the available cryogenic volume, options 2 and 3 have similar throughput (as close packing of tubes was not possible in option 3) but even if option 1 were to be expanded out to contain 37 tubes it would still have less throughput.

\textbf{Element size.}
The optical elements for options 1 and 2 are readily available. However the IR blocking filter stacks for 60~cm optics tend to involve more absorbing filters and could exceed the capacity of the pulse tubes used to cool them.  Another point worth making is that with the larger number of tubes in option 1, there is a cost, complexity, and a schedule risk of making so many elements.

\textbf{Frequency flexibility.} One advantage of smaller tube sizes is that they offer far greater flexibility in the choice of frequency bands.  In layout 3, only the center tube has sufficient optical quality in the UHF band and, even at MF, good Strehl ratios in the outer tubes are hard to obtain.  It is also far harder to deploy in stages, the smaller tubes in options 1 and 2 are easier to assemble and test so observations with a partially populated cryostat could be started far earlier.  

  \textbf{Mapping speed.} Sensitivity and systematics are by far the most important criteria.  Systematics are discussed briefly below and in far more detail in Gallardo {\it etal.}\cite{Pato10708-133}.  As mentioned, options 2 and 3 have more throughput than design 1, but also the detector spacing is more ideal.  The sensitivity calculations in Hill \etal\cite{Hill2018} shows option 2 has more sensitivity in all but the UHF bands for which option 3 has marginally more sensitivity.  Although, to fill the entire telescope focal plane, more wafers need to be fabricated in options 2 and 3, these can be added as time and budget allow.  

As a result of these points, option 1 was rejected on grounds of complexity and sensitivity. Option 3 was eliminated because of its lack of frequency flexibility and the relative difficulty in the design and manufacture of such large optics.  The choice was made to proceed with the design of cold refracting optics consisting of tubes spaced by 45~cm containing 3 lenses and 3 wafers each.

\begin{figure}
\begin{center}
  \includegraphics[height=5.5cm,angle=0]{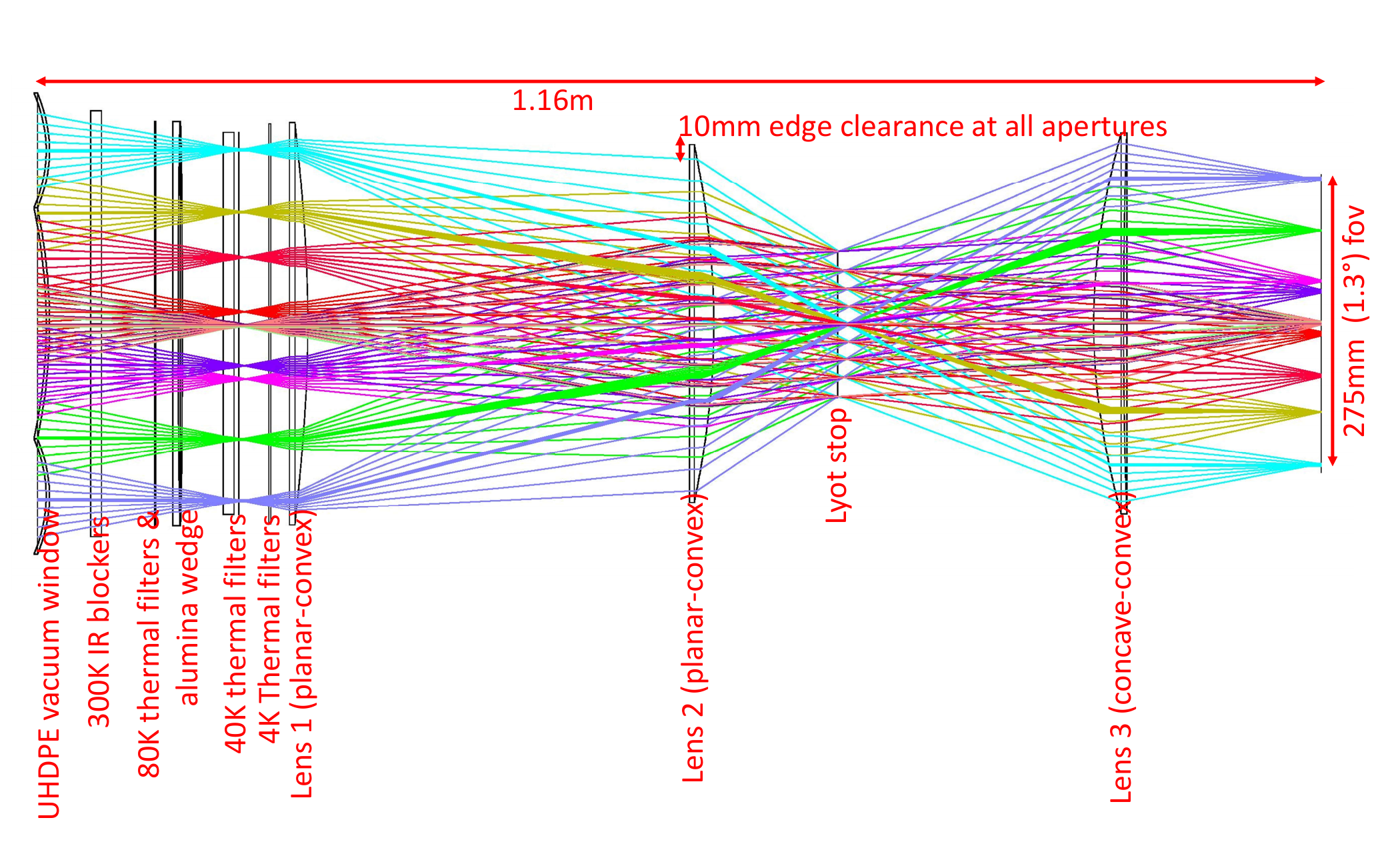}
  \includegraphics[height=6cm]{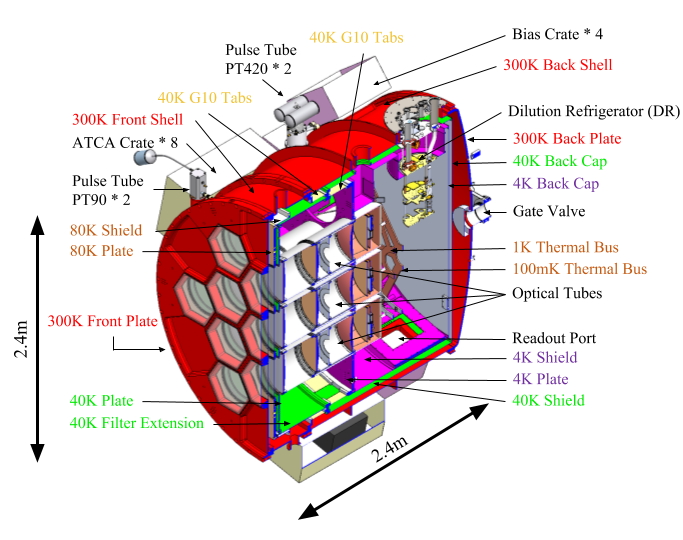}
\end{center}
\caption{\label{fig:optics}The cold optical design shown on the left as a ray-trace for a single tube and on the right as they will be installed in a cut-away view of the cryostat\cite{Zhu10708-79}.}
\end{figure}

\subsubsection{Optimization}
There are many practical constraints in the design of the optics. Finite element analysis (FEA) of the cryostat shell showed that, with a 450~mm spacing, windows much larger than $\sim$400~mm would result in an unacceptably weak cryostat front plate\cite{Orlowski-Scherer2018}. Thermal simulations showed IR blocking filters will be needed at 300~K, 80~K, 40~K and 4~K\cite{Zhu10708-79}.  With reasonable clearances and material thickness needed in this 2.4~m cryostat and room to allow extra filters (should they be needed) then $\sim$270~mm of room is needed between the front of the cryostat/vacuum window and the first 4~K elements.  Accurate placement of the filters is not needed. However, the lenses and the array need to be kept to within a few millimeters of each other (Section~\ref{sec:tol}) and so are mounted in a single tube at 4~K with 1~K and 100~mK inserts\cite{Zhu10708-79}.  To get the maximum throughput per tube, the secondary focus of the telescope is approximately halfway between the first lens and the cryostat window. For ease of construction all tubes were placed in the same plane. The telescope's focal plane is curved so the focal plane's location is set by an average of different optics tubes.

An important feature is the absorbing alumina IR filter at 80~K which is wedged for all off-axis tubes. This makes all the optics tubes in the cryostat co-axial, greatly saving on space and complexity.  Filters for the outer tubes are wedged by 1.0 degree while those on the ring are wedged by 0.6 degrees.

With this setup, the central tube was optimized using ZEMAX\footnote{Zemax LLT : https://www.zemax.com}.  As well as image quality criteria, additional constraints were added, to prevent lenses becoming larger than could be housed in 45~cm tubes, to ensure that that the edge of the 1.3 degree FoV landed on the edge of the array, that a good Lyot stop existed (using a telescope aperture of 5.5~m), that the chief rays at the array where telecentric, and that the plate scale uniform.  A number of different lens layouts were tried, but the best results were obtained using planar-convex shapes for lenses 1 and 2 and concave-convex shape for lens 3.  The best location for the Lyot stop was between lenses 2 and 3 and there was little benefit in adding more than 4 aspheric terms to each surface (although at least 3 helped).  The final optical design can be seen in Figure~\ref{fig:optics}.

After optimizing the central tube, the lens shapes were fixed and separate optimizations carried out on the inner and outer rings of optics tubes only allowing the distances between the lenses to vary.  Very little adjustment was needed.

\section{PERFORMANCE}
An analysis of this design using many criteria is described in detail in Gallardo \etal\cite{Pato10708-133}.  A brief summary is given next.
\subsection{Image quality}
\begin{figure}
\begin{center}
  \includegraphics[width=18cm]{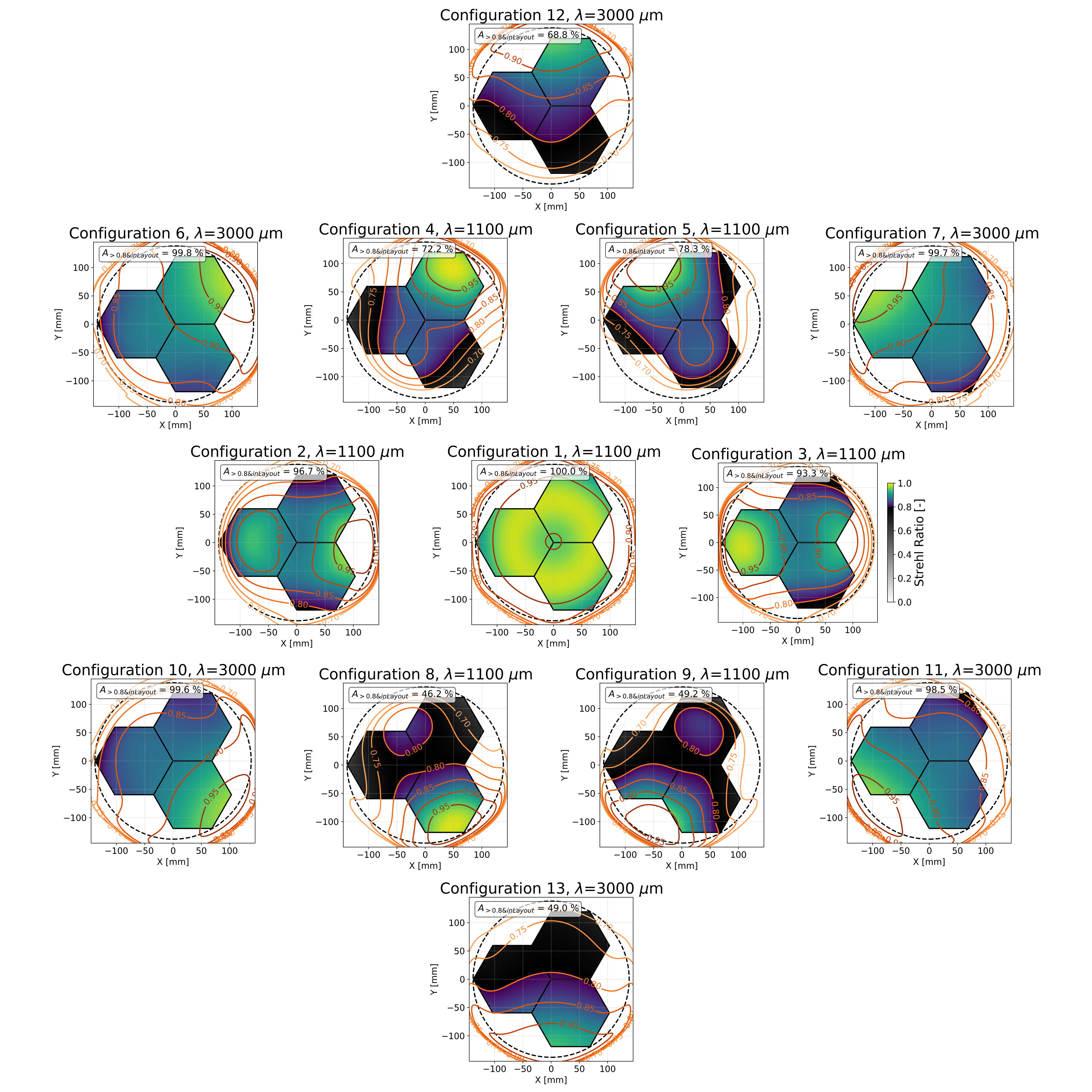}
\end{center}
  \caption{\label{fig:Strehlmaps}Strehl ratio maps for the central 7 tubes at 270~GHz and the outer 6 tubes at 100~GHz.  All 3 tubes on the middle row have more than 93\% of beams above a Strehl of 0.8 at 270~GHz. The central seven tubes at 150~GHz all have 100\% of the Strehl ratios above 0.8 with most values greater than a Strehl ratio of 0.9. }
\end{figure}

 Foremost amongst our criteria was image quality -- in our case this was measured by what fraction of the arrays in each tube had Strehl ratios above 0.8.  For our initial deployment, we require at least two of the optics tubes to be useful at UHF. As shown in Figure~\ref{fig:Strehlmaps}, three of the central tubes meet this requirement.  At least 4 of the tubes need to be used at MF and all of the seven central tubes have 100\% of the arrays above a Strehl of 0.8 at these frequencies.  Significant fractions of arrays in the outer tubes had lower optical quality but, by using different lens shapes, this could   be overcome (Section~\ref{sec:expand}).  All tubes had 100\% of their focal planes above 0.8 at LF.

\subsection{Physical optics}
Scattered light and the extra loading that can ensue has the potential to be the biggest limitation in the sensitivity of the Large SO telescope.  Using the known properties of preliminary designs for feeds and lenslets of different sizes, physical optics calculations were carried out in a reverse time sense using GRASP\footnote{TICRA software : http://www.ticra.com/software/grasp/} and ZEMAX.  The total power escaping the cold optics and missing the primary and secondary mirrors was found to be insignificant compared to our expected sky loading, even with the assumption that all this power went to ambient temperatures.

In the initial calculations all of the apertures (with the exception of the stop) were set to be 10~mm larger in radius than the lens sizes calculated using geometric optics.  Repeating the calculations with small changes in this margin were found not to affect the results so it was concluded that the clearances between the geometric rays and our apertures were large enough.  These calculations were also able to give us beam shapes that took into account the non-uniform illumination of the stop.  Where the Strehl ratios were above 0.8, the maximum ellipticities were found to be 0.2, 0.075, and 0.05 at 270~GHz,  150~GHz, and 90~GHz respectively.  With well measured beams, this was considered sufficient for our science goals.

\subsection{Ghosting}
Reflections from pairs of surfaces can create sidelobes.  Assuming worst case reflections of 100\% from the array, 1\% from silicon or Alumina surfaces, and 0.5\% from plastic windows or filters then a total ghost image from all possible pairs of reflections was created.  The peak intensity of all such images was found to be below the diffraction limit at all frequencies, so we were able to conclude that ghosting was not a significant problem.

\subsection{Tolerancing}\label{sec:tol}
No system will be built to the exact measurements specified, our knowledge of the refractive index of materials as a function of temperature is only so good and optical elements will be displaced by thermal and gravitational stress.  Some of these errors, such as the wrong coefficient of thermal expansion, can be focused out but others, such as manufacturing errors, could be different for each optics tube and must be small enough so as not to affect performance.  Inverse tolerancing was used where parameters were allowed to vary until the average Strehl decreased by 2\% or vignetting occurred. Static pointing offsets were not considered. A summary of parameters found to be most important is in table~\ref{tab:tol}. It is worth noting that these are assumed to be static tolerances --- should something change on fast timescales (such as our scan rate), far tighter limits would be needed.  The telescope and cryostat structure is expected to be rigid enough that such changes will not be a problem.
\begin{table}
\begin{center}
\begin{tabular}{|l|c|}
\hline
Relative placement of tube along optical axis wrt other tubes & +-3mm \\ \hline
Relative placement of tube perpendicular to axis wrt other tubes  & +-3mm \\ \hline
Relative tilt of tubes wrt each other & +-0.8 degrees\\ \hline 
Accuracy of angle of Alumina wedge & +-0.3 degrees \\ \hline 
Lens placement within tube (optical axis) & +-2mm \\ \hline 
Lens placement within tube (perpendicular to axis) & +-2mm \\ \hline 
Tilt of lenses within tubes & +-0.4 degrees \\ \hline 
Surface accuracy of lenses in low order Zernike modes & 0.02mm RMS \\ \hline 
\end{tabular}
\end{center}
\caption{\label{tab:tol}The tightest tolerances in our 3-lens optical design.  The values given are the smallest of any of the central 7 tubes at 270~GHz or the outer 6 tubes at 150~GHz.  Some possible variables (for example the clocking of the Alumina filters) were found to have such loose tolerances they have been omitted. For the surface accuracy of the lenses most errors are expected to be on larger scales so the quoted RMS is for the first 10 Zernike modes. }
\end{table}

As well as performance degradation from single errors, one must also consider the likely performance loss when all possible sources of error are allowed to vary together.  40 random simulations where each parameter was allowed to vary within limits in table~\ref{tab:tol} were carried out.  It was found that, even with these fairly loose tolerances, there is a 90\% chance that any decrease in Strehl will be less than 5.4\% at 270~GHz and less than 2.6\% at 150~GHz.  Given that tighter limits should be possible, it was judged that our design could be realized.

\section{FUTURE EXPANDABILITY}\label{sec:expand}
\begin{figure}
\begin{center}
\includegraphics[height=10.5cm,angle=0]{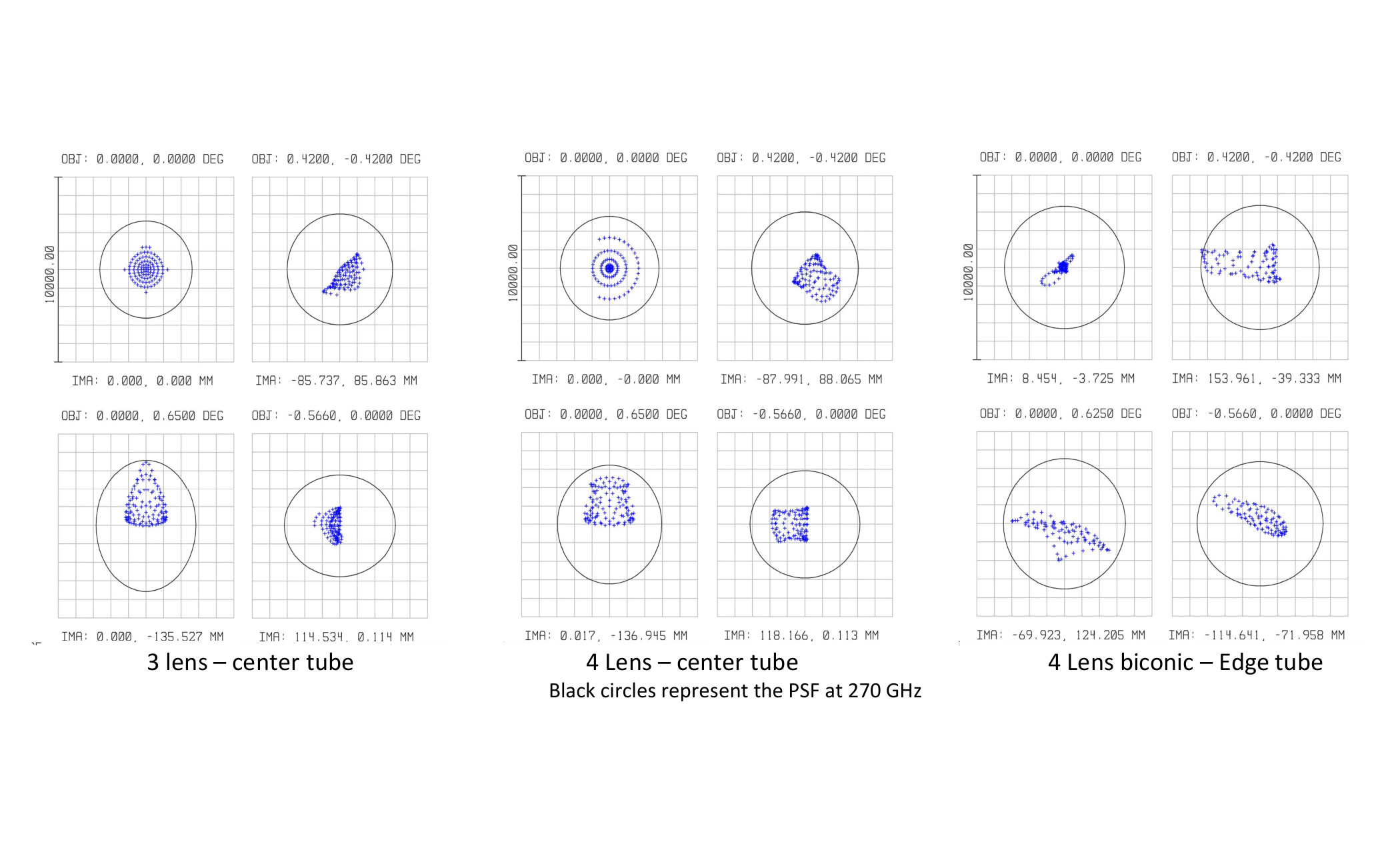} 
   \end{center}
\caption{\label{fig:expand}Possible future improvements with more complex optics systems can be seen from these spot diagrams.  On the left are the central and edge fields for the central tube of the 3 lens design presented in this paper - as can be seen variations in the plate scale make the edge beams elliptical.  The decrease in variations of plate scale of the central tube by the addition of an extra lens just above the array can be seen in the center. On the right is shown the improvement one can get by changing one of the 4 lenses to a biconic.  This is one of the outer ring of tubes, which in the 3 lens design has less than 50\% of the focal plane above a Strehl of 0.8 at 150~GHz.  With a biconic lens all Strehls are above 0.85 at 270~GHz.}
\end{figure}

Initially, the SO will deploy 7 of the 13 optical tubes. Simple upgrade paths include adding more tubes of the 3 lens design presented in this paper, or going to 4 wafers in each optics tube. However, better optical designs, were we do not restrict ourselves to 3 identical lenses, do exist.  The simplest of these is a design with 4 on-axis planar-convex lenses.  By using 4 lenses, it is possible to achieve a more uniform plate scale without sacrificing image quality at the highest frequencies and still using lenses small enough to fit within the optics tubes (Figure~\ref{fig:expand}).  Rounder beams would simplify data analysis.  

For off-axis tubes, even greater improvement can be obtained with biconic lenses.  Not only does the plate scale become more constant but outer optics tubes that can barely be used at 150~GHz with the 3 lens design become usable at 270~GHz.  Again, this can be done within the current cryostat design.  It is worth noting that no combination of tilts and decenters of symmetrical lenses was able to reproduce the improvements obtained by using biconic lenses.  Currently 3 UHF tube locations are sufficient for the SO science goals, but biconic designs could be useful in moving forwards towards CMB Stage-4. With this design, in a larger cryostat, it would be possible to house over 145\,000 90~GHz--220~GHz detectors on the SO large telescope.

\section{CONCLUSIONS}
We have presented detailed optical designs for the initial deployment on the SO Large telescope.   We found 450~mm optics tubes offered the best compromise between sensitivity, modularity and leaves us able to expand in future years.  As each optical tube has only 3 lenses and these have identical shapes for all optical tubes we have minimized cost and schedule risk while also enabling parts to be exchanged between optical tubes of the same frequency.  Detailed design of lens mounts and baffling is underway and construction should begin shortly. 

\appendix  

\section{LENS SHAPES}
\label{sec:lens_shapes}
The lens parameters for our optics.  Each surface is represented by the equation $z=\frac{c r^2}{1 + \sqrt{1 - (k+1) c^2 r^2}} + a_1 r^2 + a_2 r^4 + a_3 r^6 + a_4 r^8$. All units are in millimeters.
\subsection{lens 1}
\begin{itemize}
 \item   Clear diameter : 378.4mm
 \item   Center thickness (vertex to vertex) : 18.7mm
\end{itemize}
\begin{center}
\begin{tabular}{|l|l|c|c|c|c|c|c|}
\hline
 &	type &	 $r = 1/c$ &	$k$ &	$a_1$ &	$a_2$ &	$a_3$ &	$a_4$ \\ \hline
Surface 1  & Flat & \multicolumn{6}{|c|}{ } \\ \hline 						
Surface 2 &	Convex &-1510.61 mm & 12.0966 & -5.876941e-5 & 1.769198e-9 & -1.310183e-15 & 1.994766e-19 \\ \hline
\end{tabular}
\end{center}
\subsection{Lens 2}
\begin{itemize}
\item Clear diameter : 336 mm
\item Center thickness (vertex to vertex) : 25.8mm
\end{itemize}
\begin{center}
\begin{tabular}{|l|l|c|c|c|c|c|c|}
\hline
 &	type &	 $r = 1/c$ &	$k$ &	$a_1$ &	$a_2$ &	$a_3$ &	$a_4$ \\ \hline
Surface 1 & 	Flat & \multicolumn{6}{|c|}{ } \\ \hline 							
Surface 2 &	Convex & -871.25 mm & -10.201256 & -1.147640e-4 & 1.103643e-10 & -2.969608e-14 & 5.980271e-19 \\ \hline
\end{tabular}
\end{center}
\subsection{Lens 3}
\begin{itemize}
\item    Clear diameter : 358 mm
\item    Center thickness (vertex to vertex) : 31.55mm
    \end{itemize}
\begin{center}
\begin{tabular}{|l|l|c|c|c|c|c|c|}
\hline
 &	type &	 $r = 1/c$ &	$k$ &	$a_1$ &	$a_2$ &	$a_3$ &	$a_4$ \\ \hline
Surface 1 &	convex &	604.44 mm &	-3.177341 	& 2.366829e-5 &	-1.000417e-10 &	-6.565681e-15 &	0.0 \\ \hline
Surface 2 &	concave  & 6608.30 mm &	414.142695 &	-2.981594e-5 &	3.955202e-10 &	3.180727e-16 &	0.0 \\ \hline
\end{tabular}
\end{center}
\acknowledgments 

This work was supported in part by a grant from the Simons Foundation (Award \#457687, B.K.).  We  would also like to acknowledge the productive collaboration with the CCAT-prime team in the design of the telescope. MDN acknowledges support from NSF award AST-1454881.

\bibliography{report} 
\bibliographystyle{spiebib} 

\end{document}